\documentclass[prl,twocolumn,superscriptaddress,showpacs]{revtex4}
\usepackage{epsfig,amsmath,amssymb,graphics,color,calc}
\newcommand{\be}{\begin{equation}}
\newcommand{\ee}{\end{equation}}
\newcommand{\phii}{\phi_{\rm init}}

\begin{document}

\title{Jamming transitions in amorphous packings of frictionless spheres 
occur over a continuous range of volume fractions} 

\author{Pinaki Chaudhuri}
\affiliation{Laboratoire PMCN, 
Universit\'e Lyon 1, Universit\'e de Lyon, 
UMR CNRS 5586, 69622 Villeurbanne, France}

\author{Ludovic Berthier}
\affiliation{Laboratoire des Collo{\"\i}des, Verres
et Nanomat{\'e}riaux, UMR CNRS 5587, Universit{\'e} Montpellier 2,
34095 Montpellier, France}

\author{Srikanth Sastry}
\affiliation{Jawaharlal Nehru Centre for Advanced Scientific Research,
Jakkur Campus, Bangalore 560064, India}

\date{\today}
 
\begin{abstract}
We numerically produce fully amorphous assemblies of frictionless
spheres in three dimensions and study the jamming transition these
packings undergo at large volume fractions.  We specify four protocols
yielding a critical value for the jamming volume fraction which is
sharply defined in the limit of large system size, but is different
for each protocol. Thus, we directly establish the existence of a
continuous range of volume fraction where nonequilibrium jamming
transitions occur. However, these jamming transitions share the same
critical behaviour.  Our results suggest that, even in the absence of
partial crystalline ordering, a unique location of a random close
packing does not exist, and that volume fraction alone is not
sufficient to describe the properties of jammed states.
\end{abstract}

\pacs{05.10.-a, 05.20.Jj, 64.70.Pf}

\maketitle

Everyday experience shows that it is not possible
to compress a disordered assembly of rigid particles beyond a
maximal packing fraction. The idea of a critical value
for the volume fraction of this `random close packing' 
has a long history in science~\cite{bernal}. 
For the idealized case of hard, spherical, 
frictionless particles the idea recently emerged that 
a critical value of the volume fraction, named $\phi_J$, can be 
appropriately defined in the thermodynamic limit~\cite{pointJ,makse}, 
and that it corresponds to a critical 
point (`point $J$') with remarkable scaling properties observed when $\phi_J$ 
is approached from either side~\cite{pointJ,epitome}. A unique critical 
packing fraction is also found under quasi-static~\cite{claus} or 
continuous~\cite{teitel,hatano} shearing.
Further studies have also established the peculiar properties
of configurations prepared exactly at $\phi_J$, 
with specific, singular behaviour of the pair correlation 
function~\cite{donev,leo}. This leads to the conjecture that
the properties of this critical point can influence the physical behaviour 
of dense particle systems under various conditions~\cite{liunagel}.
    
The identification of a unique critical point for jamming 
is surprising because two distinct phenomena occur when compressing 
hard spheres at equilibrium. First, an equilibrated
system of monodisperse  hard spheres 
is fully crystalline above $\phi=0.54$. Since crystalline states
are efficiently packed, it is in principle always possible to 
increase the volume fraction of a disordered assembly by increasing the 
local order~\cite{torquato2}. 
In the studies mentioned above, crystallization is 
prevented by using size polydispersity which increases 
the nucleation barrier and can efficiently suppress crystallization at
equilibrium. Moreover, these studies employ athermal numerical 
protocols which do not allow the system to reach equilibrium.

A second relevant phenomenon occurring at equilibrium is the 
glass transition. 
Compressing hard spheres at thermal equilibrium, 
it is found that ergodicity is lost near a volume fraction 
much lower than $\phi_J$, because the relaxation time for structural
relaxation becomes larger than experimental timescales~\cite{luca}.  
Necessarily, then, the properties of hard sphere glasses
should depend on the preparation history, as is 
well-known for molecular glasses~\cite{sastrynat98}. 
The resulting interplay between 
glass and jamming transitions is currently receiving attention, both
from theory and simulations. A special density akin to $\phi_J$ 
appears in mean-field glasses undergoing 
a random first order transition~\cite{jorge}. 
In this language, $\phi_J$ corresponds 
to the (sharp) appearance of an exponentially large number of metastable
states, and such states exist along a continuous range
of densities between $\phi_J$ and that of the  `glass close 
packing'~\cite{jorge,zamponi}, 
where the configurational entropy (or complexity)
counting metastable states vanishes. However, none of these mean-field 
concepts is expected to remain sharply defined in three 
dimensions~\cite{zamponi}, and
it is important to assess the validity of this interpretation
in finite dimensions.

Previous numerical work has shown that different compression rates
produce glasses with a pressure which appears to diverge at different
densities, at least for finite size
systems~\cite{speedy1,speedy2,torquato,hermes}.  Indeed, the analogy
\cite{speedy3,reich} of the jamming density with the energy of
inherent structures in systems with soft potentials
\cite{sastrynat98}, suggests that the jamming densities should depend
both on compression rates as well as the initial state from which the
glasses are generated. Recently, a compression protocol
was specifically devised to prevent any ordering, thus allowing to
focus directly on the sole influence of glassy behaviour on
jamming~\cite{tom}. Starting from a well-equilibrated
fluid configuration at a given volume fraction, $\phii$, one uses
a very large compression rate to reach nearly jammed
configurations. During these rapid compressions, the glass pressure
was observed to diverge within a finite range of volume fractions,
even in the thermodynamic limit~\cite{tom}, but the jamming transition
was not studied.

Since evidence is mounting in favour of both a unique location
of the jamming transition with critical properties 
in some papers, or a continuous range of volume fraction
in some other works, it appears timely to reconcile these two lines
of research and answer the following important, open questions. Does 
a continuous range of volume fraction for jamming necessarily 
result from crystallization or demixing? Can one reconcile 
the results found using thermal and athermal protocols?
Can the remarkable properties of the jamming transition
survive if its location is not unique?
Here, we provide precise answers to these questions and 
close the gap between two sets of ideas.
We study numerically how amorphous assemblies 
of hard spheres prepared using the (thermal) tools of Ref.~\cite{tom}
jam at large volume fraction. We then apply the alternative 
(athermal) tools  
of Ref.~\cite{pointJ} to analyze the properties of jammed configuration
at and near the transition. For the specific three-dimensional 
binary system we use, we can directly 
establish the existence of a continuous line of $J$-points 
extending at least in the range 
$\phi_J \in [0.648, 0.662]$, all sharing similar 
critical properties 
and exponents. Thus, although point $J$ is not unique, 
its critical properties are.

We study the jamming transition in a 
three-dimensional system of spherical particles using periodic boundary 
conditions. We use a
50:50 binary mixture of spheres with diameter ratio 
$1.4$. 
Our preparation protocol combines the fast compression of hard 
spheres described above~\cite{tom}, to the 
athermal compression of spheres treated as soft spheres 
as in Ref.~\cite{pointJ}.
Specifically, the particles are first treated as hard spheres
and equilibrated using Monte-Carlo simulations over a broad
range of volume fraction up to $\phi=0.596$.
This allows one to obtain disordered configurations 
representative of the metastable fluid
of hard spheres (no crystallisation or demixing is observed for simulations
as long as $10^{10}$ Monte-Carlo timesteps). We then use 
Monte-Carlo simulations to rapidly compress these 
equilibrated fluid configurations. During these compressions
the system has no time to relax and retains a structure very 
close to the initial fluid states. 
Thus, contrary to previous work employing 
slow compressions~\cite{torquato,hermes},
we are certain that the jammed configurations
we produce contain no more order than the original 
equilibrated fluid configurations. 

\begin{table}
\begin{tabular}{|c | c c c c|}
\hline
$\phii$ & 0.3572 & 0.5397  &  0.5672 &  0.5935 \\
\hline 
$\phi_J$ ($8000,\phii$) & {\bf 0.6481} & {\bf 0.6499 } & {\bf 0.6537} & 
{\bf 0.6616} \\
$\phi_J$ ($1000,\phii$) & 0.6466 & 0.6491 & 0.6531 & 0.6616 \\
\hline
$\sigma_J$ ($8000,\phii$) & $5.2  \cdot 10^{-4}$ & $4.9  \cdot 10^{-4}$ & 
$3.0  \cdot 10^{-4}$ & 
$2.5  \cdot 10^{-4}$ \\
$\sigma_J$ ($1000,\phii$) & $9.7 \cdot 10^{-4}$ & $1.0 \cdot 10^{-3}$ & 
$7.6  \cdot 10^{-4}$ & 
$5.8  \cdot 10^{-4}$ \\
\hline
$z$ at $\phi_J$ & 6.0017 & 6.0022 & 6.0021 &  6.0023 \\
$\sigma_z$ & 2.47 & 2.50 & 2.54 & 2.65 \\
$z_{12}$ at $\phi_J$ & 2.92 & 2.91 & 2.92 & 2.97  \\
\hline
Rattlers & 4.3 \% & 4.7 \% & 5.4 \% & 6.2 \% \\
\hline
\end{tabular}
\caption{\label{table} Statistics of jammed configurations 
for different system sizes $N$ and various
$\phii$. The limit $\phi_J(N \to \infty,\phii)$ exists, 
with finite $N$ fluctuations decaying 
in good agreement with $N^{-1/2}$, and retains a dependence on $\phii$.
At the transition, configurations are all 
isostatic, $z \approx 6$ (with similar distribution width $\sigma_z$), 
nearly identical local structure, number of rattlers, and
no increasing demixing between species ($z_{12}$ is nearly constant).}
\end{table}

When (reduced) pressure is very large, $P/(\rho k_B T) \sim10^3$, 
we stop the compression and switch to the athermal procedure of 
Ref.~\cite{pointJ}, where spheres are now treated as 
harmonic spheres interacting through a soft pair potential,
$V_{\rm soft}(r_{ij}) = (1-r_{ij}/\sigma_{ij})^2$. Here, $\sigma_{ij}
= (\sigma_i+\sigma_j)/2$ and $\sigma_i$ is the radius of particle $i$.
The compression then 
proceeds in a succession of small instantaneous particle inflation 
followed by energy minimization using conjugate gradient~\cite{pointJ}. 
At large volume fraction, the energy 
cannot be minimized to zero anymore, and particles now overlap, signalling
that the jamming transition has been crossed.  
We emphasize that all the details entering this 
compression protocol in principle 
quantitatively affect the results presented 
below~\cite{donev2}. However, since the most sensitive control
parameter is $\phii$, we shall only use its value to 
distinguish between the different protocols, all other parameters
being kept fixed. 
We focus on four values for $\phii$ listed
in Table~\ref{table}. We monitor finite size effects
and convergence towards the thermodynamic limit by studying
two system sizes, $N=1000$ and $8000$, chosen in the regime
where scaling with $N$  is well understood~\cite{pointJ}.

\begin{figure}
\includegraphics[height=4.2cm,angle=-90,clip]{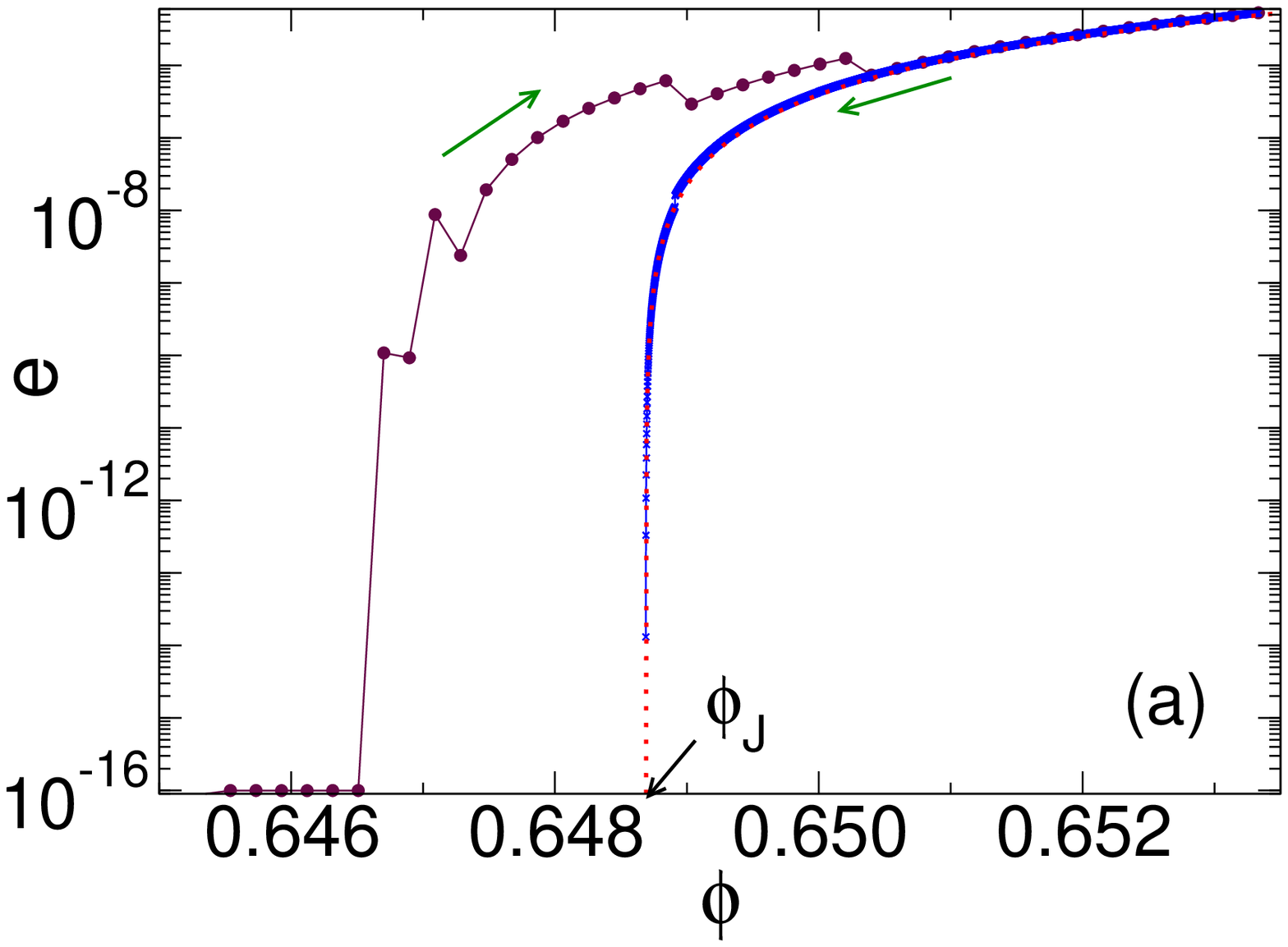}
\includegraphics[height=4.2cm,angle=-90,clip]{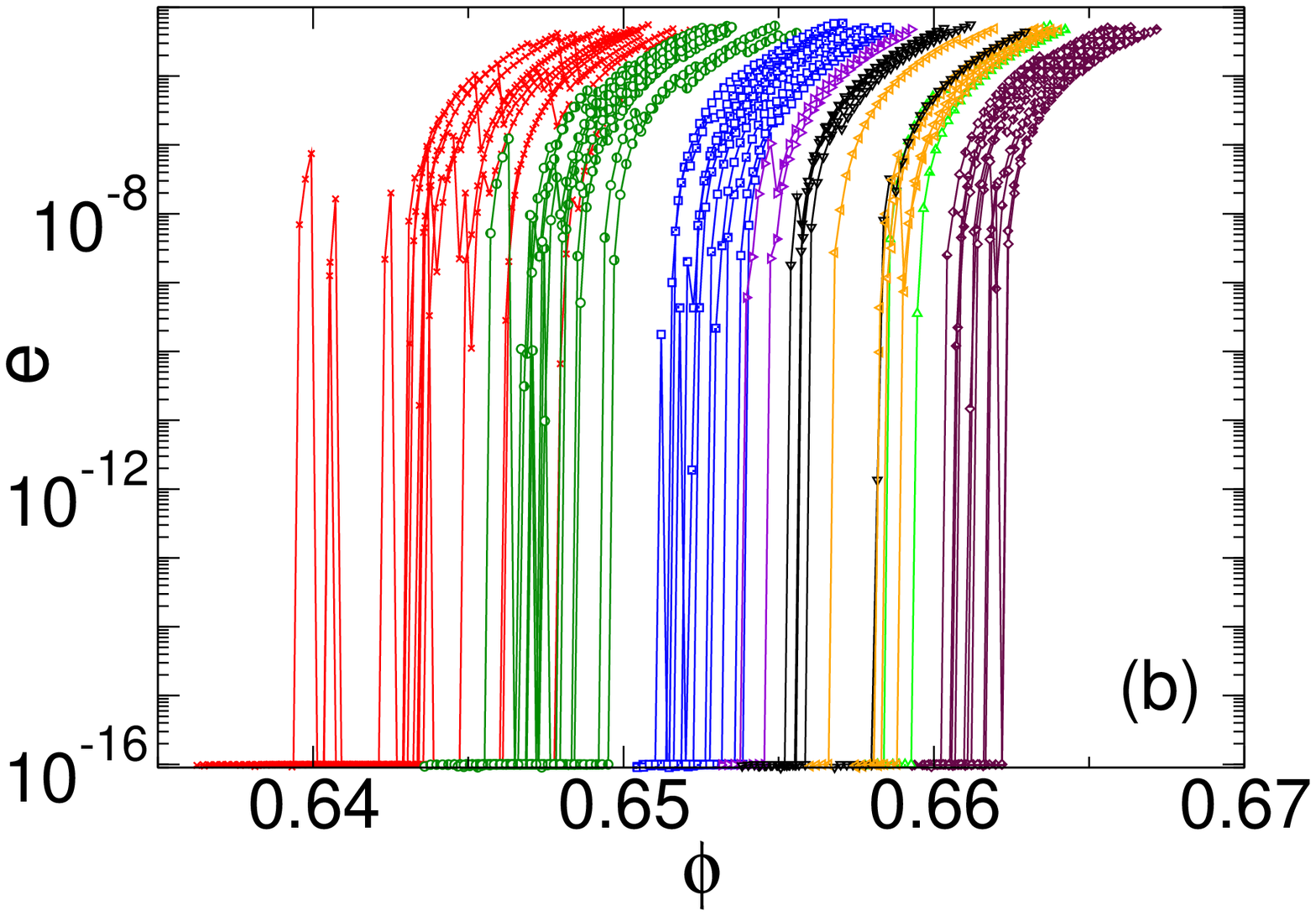}
\includegraphics[height=4.2cm,angle=-90,clip]{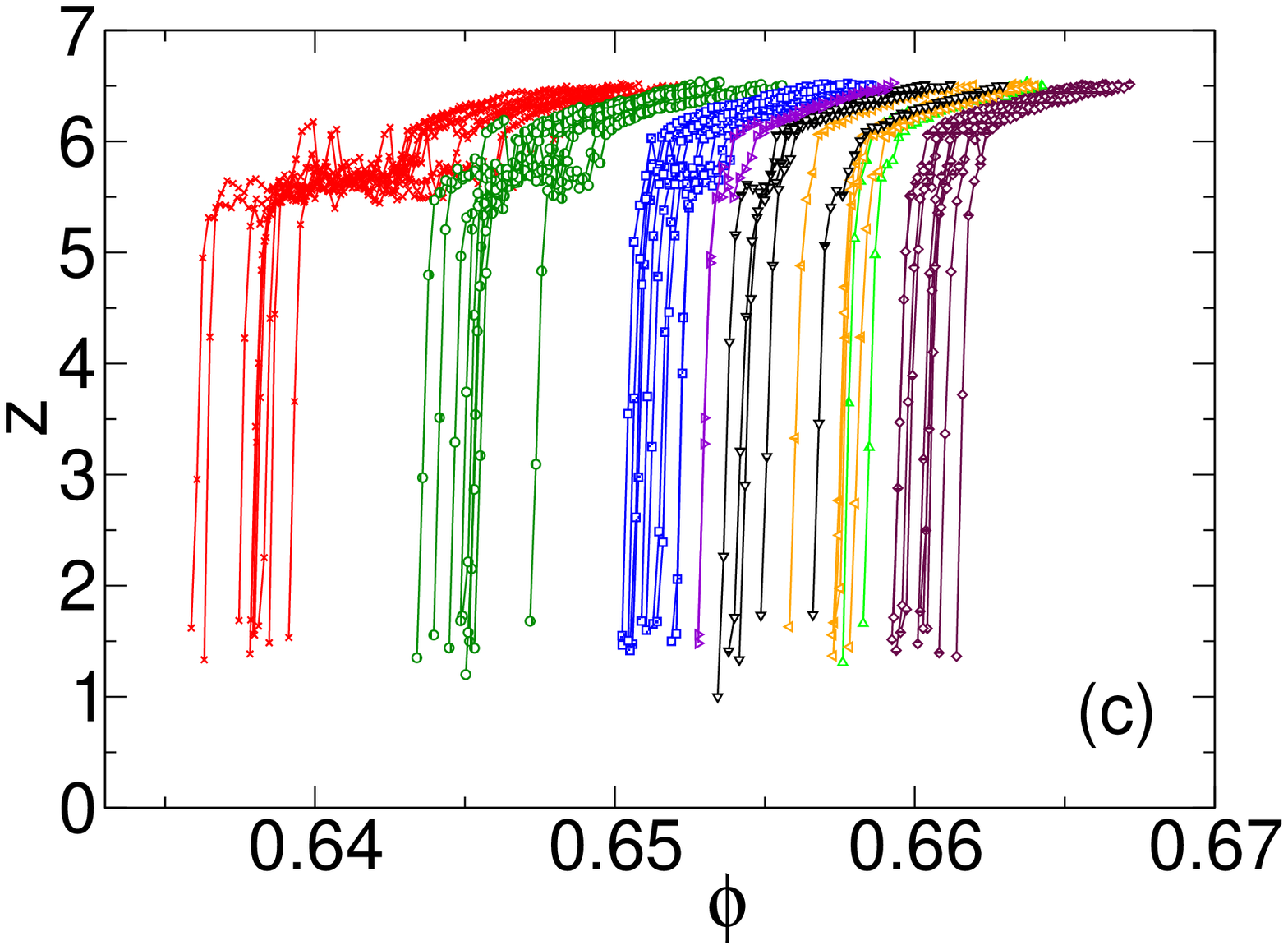}
\includegraphics[height=4.2cm,angle=-90,clip]{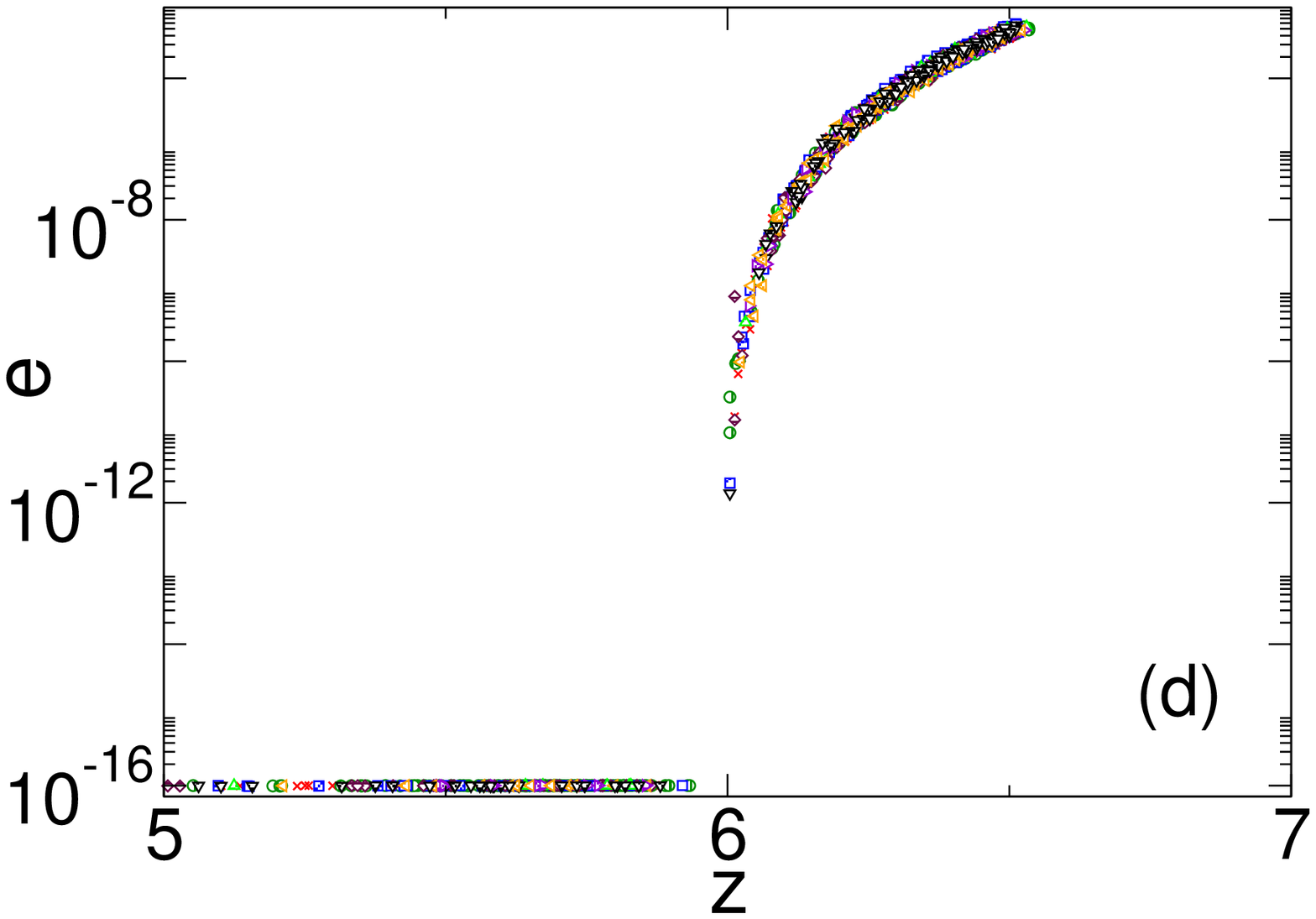}
\caption{\label{fig1} 
(a) The final part of a compression-decompression cycle across the jamming
transition for $N=10^3$ and $\phii=0.3572$, showing the practical
determination of $\phi_J$ for this sample.
(b) 
Evolution of the energy density during 
compressions of several runs with $N=1000$. Different 
curves with the same color correspond to independent 
realizations at constant $\phii$, while 
different $\phii$ are shown with different colors. 
(c) Same as (b) for the evolution of the number of contacts
per particle.
(d) All data in (b) and (c) collapse when 
$\phi$ is eliminated, showing that the jamming transition 
is better defined by its properties than by its location.}
\end{figure}

In Fig.~\ref{fig1}-a we show the final part 
(e.g. when particles are treated as soft) of a randomly selected 
compression history for a 
system with $N=10^3$. We follow the 
energy density of the system, $e = N^{-1} \sum_{i<j} V_{\rm soft}(r_{ij})$,
for increasing $\phi$. 
We have defined configurations having energy density less 
than $10^{-16}$ as unjammed.
For low enough $\phi$, we obtain only unjammed particle configurations.
Upon increasing the density, there appears a volume 
fraction above which the energy increases rapidly above
zero: this is the jamming transition~\cite{pointJ}. Upon further compression, 
the energy keeps increasing on average, and so does the 
average number of contacts per particle, $z$. 

This qualitative description
becomes more complicated when multiple histories are superimposed,
see Fig.~\ref{fig1}-b. There are several sources of fluctuations 
in this plot. 
A first source of fluctuations stems from the possibility for the system
to undergo some reorganization 
which makes the energy drop suddenly, and makes the
$e(\phi)$ curve multivalued during a single compression run.
Thus, no simple scaling law for the energy can be detected
during compressions.
The second source of fluctuations arises when different 
realizations of the same protocol (i.e. same value 
of $\phii$ but starting from independent fluid 
configurations) are compared. 
These fluctuations 
are present because $N$ is finite, and should vanish when 
$N \to \infty$. 
A third source of fluctuations can be seen in Fig.~\ref{fig1} when
$\phii$ is varied, larger $\phii$ typically yielding
larger volume fraction for the onset of positive energies. 

If we decompress jammed configurations, reorganizations are much 
less likely, and the $e(\phi)$ curve is smooth enough that 
it allows a quantitative determination of $\phi_J$, see
Fig.~\ref{fig1}-a.
Of course, $\phi_J$ now depends on one additional parameter, 
namely the volume fraction from which 
decompression starts. We arbitrarily decompress when $e \approx
10^{-7}$ and obtain $\phi_J$ by fitting $e$ to a power 
law decay, $e \sim (\phi-\phi_J)^2$
during decompression. Thus, we determine $\phi_J = \phi_J(N,\phii)$.
We repeat this analysis for different $N$, $\phii$, and
initial configurations to obtain the statistics of 
$\phi_J$ reported in Table~\ref{table}.
The main result of this analysis 
is that $\phi_J(N \to \infty, \phii)$ seems sharply 
defined (the fluctuations decrease in good agreement with the 
$N^{-1/2}$ scaling~\cite{pointJ}) for each value 
of $\phii$, but the $\phii$ dependence 
survives a substantial increase of system size. 
Therefore, we have directly established 
that, for frictionless spheres, 
the jamming transition occurs along a {\it continuous line} 
rather than at a specific point in volume fraction, as hinted in earlier
simulations~\cite{hermes,tom}, and predicted 
theoretically~\cite{jorge,zamponi}.

\begin{figure}
\includegraphics[width=7.5cm,angle=0,clip]{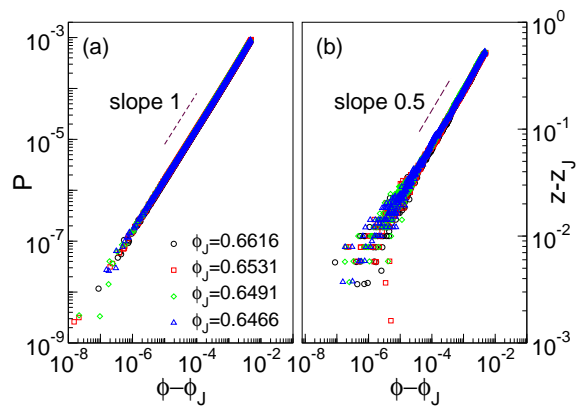}
\caption{\label{fig2} Scaling behaviour along the 
continuous line of $J$ points.
The pressure vanishes as $P \sim (\phi-\phi_J)$ (left axis) while
the number of contacts (right axis) is discontinuous, 
$z \sim z_c + \sqrt{\phi -\phi_J}$ with $z_c \simeq 6$. 
We show data for 4 representative points $J$ with 
4 different $\phi_J$.}
\end{figure}

Having found several critical points for jamming, rather than one,
we now ask whether these different $J$-points are equivalent.
We have measured the static structure 
of the jammed configurations through the (partial) pair correlation functions 
$g_{\alpha \beta}(r)$ ($\alpha, \beta = 1, 2$ 
for small/large particles, respectively) 
and the statistics of contacts between particles.
We have also followed the properties of configurations 
approaching jamming from large volume fraction.
As in~\cite{pointJ}, we find several scaling laws near our 
four $J$-points, for $P$, $e$ and $z$. 
In Fig.~\ref{fig2} we show that for all $J$-points we have
$P \sim (\phi-\phi_J)$
and $z=z_c +\sqrt{\phi-\phi_J}$
with $z_c \simeq 6$ (see Table~\ref{table}).
Within our numerical precision, 
the critical properties of jamming transitions along the 
$J$-line are identical.

This analysis suggests that the value of $\phi_J$ is strongly influenced
by the entire sample history (compression/decompression protocol, 
reorganization during compressions, etc.), but that the 
scaling properties of the jamming transition are much more robust.
It is tempting to formulate the appealing 
hypothesis that the history dependence of the jammed configurations 
are contained solely in the volume fraction $\phi_J$ at which they jam. 
This would suggest that data analysis of the jamming transition 
should drastically simplify if we eliminate $\phi_J$ from consideration. 
We confirm the validity of our hypothesis in Fig.~\ref{fig1}-d, where 
the three sources of fluctuations discussed above are accounted 
for by eliminating $\phi$ from the description. 
By plotting $e$ vs. $z$, we find that data for all cases nicely collapse
onto a mastercurve, $e \sim (z-z_c)^{4}$, independently 
of the sample preparation history. Thus, the details of the 
numerical protocols are irrelevant to the relationship between energy 
and coordination number across the jamming transition. 

\begin{figure}
\includegraphics[width=8.5cm,angle=-0,clip]{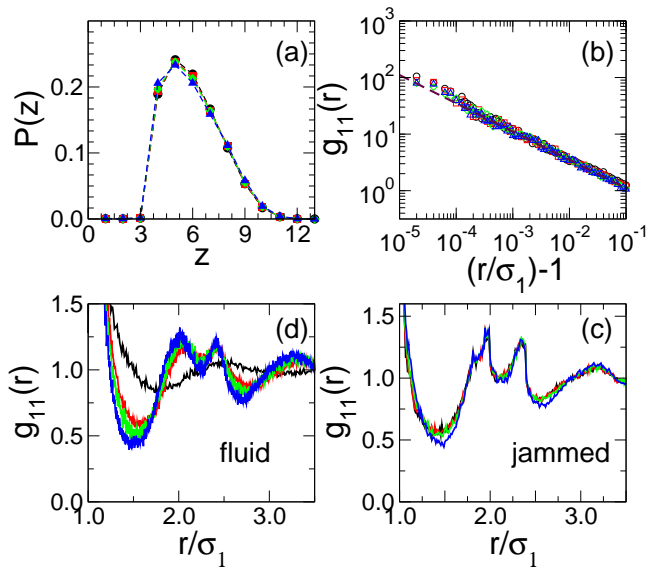}
\caption{\label{fig3} 
(a) Statistical distribution of the number of 
contacts per particle, $z$, for 4 points $J$.
(b) Power law singularity near contact for $g_{11}(r)\sim 
(r-\sigma_{11})^{\gamma}$ for the four points $J$, with $\gamma = \frac{1}{2}$ shown as dashed line.
(c) $g_{11}(r)$ near second and third peaks for 4 points $J$.
(d) As in (c) for the equilibrated fluid configurations at $\phii$.}
\end{figure}   

Not only are scaling properties near jamming robust,
but the special structure of the pair correlation function at the
transition is also observed in all our samples.  The pair correlation
has a delta peak at contact, which, when integrated yields the average
number of contacts, see Table~\ref{table}. The total number, $z$, and
partial numbers, $z_{\alpha \beta}$, have the same value for all
$J$-points, within numerical accuracy. Note in particular that
$z_{12}$ does not evolve, indicating that demixing between species of
the mixture is not observed.  In fact the full distributions of
contact number barely change, see Fig.~\ref{fig3}-a. The pair
correlation also has a power law singularity near
contact~\cite{donev,leo}, $g_{\alpha \beta} (r) \sim (r-\sigma_{\alpha
  \beta})^{\gamma}$. This is shown in Fig.~\ref{fig3}-b for
$g_{11}(r)$, where again the exponent is the same for the four
$J$-points, $\gamma \approx 0.5$. In Fig.~\ref{fig3}-c, we observe a
subtle but systematic evolution with $\phii$ of $g(r)$ 
from near the first minimum to the third peak. Similar to the case of
inherent structures in systems with soft potentials, these subtle variations
reflect the more clearly visible changes in the structure of the
equilibrated fluid configurations at $\phii$ which were used as
starting points for the compressions, as shown in Fig.~\ref{fig3}-d.
Finally, we note that the increase of $\phi_J$ does not result simply
from a decreasing number of rattlers (particles with no contact), see
Table~\ref{table}.

Our results directly establish that a unique
location for the jamming transition
cannot be expected, even when crystallization or demixing play no
role. Rather, our findings are broadly consistent with a `landscape' picture
wherein the fluid explores a phase space that does not form the basin
of a well-defined, unique ground state, but instead one marked by the
presence of many local minima or metastable states, as is typical of
glassy systems. The basins predominantly sampled vary with the
equilibrium volume fraction of the system. As in thermal 
systems~\cite{sciortino} using the inherent structure formalism, the nature,
number and evolution with volume fraction of these metastable states
can be evaluated~\cite{speedy3}, and are described exactly in the
mean-field limit~\cite{jorge,zamponi}. An evaluation of
configurational entropy associated with these jammed states
and a comparison with the sharp boundaries predicted in mean-field 
theory, corresponding to the appearance and vanishing of a
configurational entropy, would be valuable for this system.

In summary, we have shown that jamming transitions
of amorphous packings of frictionless spheres occur 
along a continuous range of volume fraction. These transition
are sharply defined in the thermodynamic limit, and this finite
range exists even when ordering phenomena 
or friction are absent, contrasting with
the idea of a  unique jamming point.
Since the location of the jamming transition in fact results 
from the specific protocol used to study it, 
we see no reason why it should be reproducible
from one experiment to another. 

We would like to thank J.-L. Barrat and J. Kurchan for useful
exchanges. PC acknowledges financial support from  
UCBL1 and ANR SYSCOMM.

\end{document}